\documentclass[english,11pt,a4paper]{article}
\usepackage[T1]{fontenc}
\usepackage[latin9]{inputenc}
\usepackage{amsmath}
\usepackage{amsthm}
\usepackage{amssymb}
\usepackage{graphicx}

\makeatletter
\numberwithin{equation}{section}

\usepackage{jheppub}
\usepackage{slashed}

\makeatother

\usepackage{babel}
\begin{document}
\title{ The $\nu_{R}$-philic scalar dark matter}

\abstract{ Right-handed neutrinos ($\nu_{R}$) offer an intriguing
portal to new physics in hidden sectors where dark matter (DM) may
reside. In this work, we delve into the simplest hidden sector involving
only a real scalar exclusively coupled to $\nu_{R}$, referred to
as the $\nu_{R}$-philic scalar. We investigate the viability of the
$\nu_{R}$-philic scalar to serve as a DM candidate, under the constraint
that the coupling of $\nu_{R}$ to the standard model is determined
by the seesaw relation and is responsible for the observed DM abundance.
By analyzing the DM decay channels and solving Boltzmann equations,
we identify the viable parameter space. In particular, our study reveals
a lower bound ($2.6\times10^{5}$ GeV) on the mass of $\nu_{R}$ for
the $\nu_{R}$-philic scalar to be DM. The DM mass may vary from sub-MeV
to sub-GeV. Within the viable parameter space, monochromatic neutrino
lines from DM decay can be an important signal for DM indirect detection.

}

\author[a]{Xun-Jie Xu,} 
\author[b]{Siyu Zhou,} 
\author[a,c]{Junyu Zhu} 
\affiliation[a]{Institute of High Energy Physics, Chinese Academy of Sciences, Beijing 100049, China} 
\affiliation[b]{University of Science and Technology of China, Hefei, Anhui 230026, China} 
\affiliation[c]{School of Physical Sciences, University of Chinese Academy of Sciences, Beijing 100049, China}
\preprint{\today}  
\emailAdd{xuxj@ihep.ac.cn} 
\emailAdd{zhouker@mail.ustc.edu.cn} 
\emailAdd{zhujunyu@ihep.ac.cn} 
  
\maketitle

\section{Introduction}

Right-handed neutrinos ($\nu_{R}$) are among the most compelling
extensions of the Standard Model (SM), offering a simple solution
to the problems of neutrino mass and matter-antimatter asymmetry.
   Given that the observed dark matter (DM) in our universe is
also evidence of new physics beyond the SM, it is tempting to consider
potential connections between $\nu_{R}$ and DM.    

Being nearly invisible,   right-handed neutrinos themselves have
long been considered as a popular DM candidate, known as keV sterile
neutrino DM---see Refs.~\cite{Dasgupta:2021ies,Abazajian:2017tcc,Drewes:2016upu}
for recent reviews. However, the simplest scenario achieved via the
Dodelson--Widrow mechanism~\cite{Dodelson:1993je} has been ruled
out by X-ray and Lyman-$\alpha$ observations~\cite{Dasgupta:2021ies,Palanque-Delabrouille:2019iyz}. 

Due to their singlet nature, right-handed neutrinos are also considered
as a portal to more hidden sectors, where a DM candidate may reside.
For instance, one can introduce a dark fermion-scalar pair, $\chi\text{-}\phi$,
with the interaction ${\cal L}\supset\nu_{R}\chi\phi$. This is the
minimal framework to accommodate absolutely stable DM in the $\nu_{R}$
sector and has  been investigated extensively in the literature \cite{Pospelov:2007mp,Falkowski:2009yz,Gonzalez-Macias:2016vxy,Escudero:2016ksa,Tang:2016sib,Batell:2017rol,Batell:2017cmf,Bandyopadhyay:2018qcv,Becker:2018rve,Chianese:2018dsz,Folgado:2018qlv,Yin:2018yjn,Chianese:2019epo,Hall:2019rld,Blennow:2019fhy,Bandyopadhyay:2020qpn,Hall:2021zsk,Chianese:2021toe,Biswas:2021kio,Coy:2021sse,Coy:2022xfj,Barman:2022scg,Coito:2022kif,Bandyopadhyay:2022xlp,Li:2022bpp,Ghosh:2023ocl,Ahmed:2023vdb}.
Within this framework, the relic abundance of DM depends not only
on the coupling of $\nu_{R}\chi\phi$, but also on the coupling of
$\nu_{R}$ with the SM. If the $\nu_{R}$-SM coupling is sufficiently
weak while $\nu_{R}$ dominantly decays to $\chi$ and $\phi$, the
DM abundance depends only on the $\nu_{R}$-SM coupling~\cite{Coy:2021sse,Coy:2022xfj}.
Assuming the $\nu_{R}$-SM coupling is related to neutrino masses
via the type-I seesaw relation, the DM abundance largely depends on
the $\nu_{R}$-DM coupling since the type-I seesaw Yukawa coupling
can maintain thermal equilibrium between $\nu_{R}$ and the SM~\cite{Barman:2022scg}.
 A comprehensive study that systematically investigates all possible
scenarios can be found in Ref.~\cite{Li:2022bpp}.

In this work, we consider a further simplified framework in which
the dark sector contains only a scalar singlet $\phi$ and it is exclusively
coupled to $\nu_{R}$~\cite{Dudas:2014bca}. We refer to it as the
$\nu_{R}$-philic scalar. Despite its simplicity, the $\nu_{R}$-philic
scalar as a DM candidate has drawn less attention since it lacks
absolute stability.  Nevertheless, the absolute stability is not
fundamentally necessary for a DM candidate. It is possible, with a
small coupling, that the $\nu_{R}$-philic scalar can have a sufficiently
long lifetime\footnote{If it is short-lived at the cosmological timescale and cannot be DM,
there are still rich cosmological implications---see e.g.~\cite{Li:2023kuz}
for a dedicated study on the Majoron in the early universe. }. Therefore, we are curious about whether the $\nu_{R}$-philic scalar,
as the minimal dark sector  built on $\nu_{R}$, can be a DM candidate
or not. 

In this paper, we present a dedicated analysis of possible decay channels
of the $\nu_{R}$-philic scalar $\phi$, solve the Boltzmann equation
of $\phi$, and identify the viable parameter space for it to be DM.
 We find that, to serve as a DM candidate, the $\nu_{R}$-philic
scalar needs to be lighter than $\sim200$ MeV while the $\nu_{R}$
mass needs to be above $\sim10^{4}$ GeV, provided that the $\nu_{R}$-SM
coupling is determined by the seesaw relation.  

The $\nu_{R}$-philic scalar DM shares some similarities with the
Majoron DM~\cite{Berezinsky:1993fm,Bazzocchi:2008fh,Gu:2010ys,Frigerio:2011in,Lattanzi:2013uza,Queiroz:2014yna,Boucenna:2014uma,Garcia-Cely:2017oco,Reig:2019sok,Akita:2023qiz}.
The Majoron, with the name implying its relevance to the generation
of the Majorana mass of $\nu_{R}$, is primarily coupled to $\nu_{R}$
and could be included in our $\nu_{R}$-philic scalar framework. However,
as a Goldstone boson arising from the lepton number symmetry breaking,
the Majoron would be massless in the minimal construction~\cite{Chikashige:1980ui}.
To be DM, it has to possess a mass, which is typically introduced
via soft-breaking terms. For example, in Ref.~\cite{Frigerio:2011in},
a Majoron-Higgs quartic term was introduced to simultaneously account
for both the Majoron mass and the relic abundance. In the presence
of such additional interactions, our conclusion in general does not
apply but our analysis can be considered as a case study in the vanishing
limit of additional interactions. 

Our work is structured as follows. In Sec.~\ref{sec:Basic}, we introduce
the model considered in this work as well as some basic formulae for
our analysis. In Sec.~\ref{sec:Viability}, we discuss the dominant
decay channels of the $\nu_{R}$-philic scalar, compute the decay
widths, and estimate the thermal production of DM in the early universe.
This allows us to identify the viable parameter space in an almost
analytical approach, while the results are further supported by numerically
 solving the Boltzmann equations in Sec.~\ref{sec:Solving}. In Sec.~\ref{sec:Neutrino-signals},
we discuss possible neutrino signals from DM decay as an observational
consequence of this model. Finally we conclude in Sec.~\ref{sec:Conclusion}
and relegate some detailed calculations to the appendix.

\section{Basic setup\label{sec:Basic}}

\subsection{The Lagrangian}

We consider the standard Type-I seesaw model extended by a scalar
singlet. The full Lagrangian of the model considered in this work
reads:
\begin{equation}
{\cal L}={\cal L}_{{\rm SM}}+{\cal L}_{{\rm seesaw}}+\frac{1}{2}(\partial\phi)^{2}-\frac{1}{2}m_{\phi}^{2}\phi^{2}+\left(\frac{y}{2}\thinspace\nu_{R}\nu_{R}\phi+{\rm h.c.}\right),\label{eq:L}
\end{equation}
where $\phi$ is a real scalar with mass $m_{\phi}$ and couples only
to  $\nu_{R}.$ The SM Lagrangian is denoted by ${\cal L}_{{\rm SM}}$
and the seesaw Lagrangian is given by 
\begin{equation}
{\cal L}_{{\rm seesaw}}=\nu_{R}^{\dagger}\overline{\sigma}^{\mu}i\partial_{\mu}\nu_{R}-\left(\frac{m_{R}}{2}\nu_{R}\nu_{R}+{\rm h.c.}\right)+\left(y_{\nu}\tilde{H}^{\dagger}L\nu_{R}+{\rm h.c.}\right),\label{eq:}
\end{equation}
where $\tilde{H}\equiv i\sigma_{2}H^{*}$ with $H$ the SM Higgs doublet,
 and $L=\left(\nu_{L},\ e_{L}\right)^{T}$ is the lepton doublet.
In this work, for simplicity, we do not consider the flavor structure
within the neutrino sector. Consequently,  we treat $y$, $m_{R}$,
and $y_{\nu}$ as single values rather than matrices.  Additionally,
we adopt the notation of two-component {\it Weyl-van der Waerden}
spinors to denote fermions ($\nu_{R}$, $\nu_{L}$, and $e_{L}$),
following the convention reviewed in Ref.~\cite{Dreiner:2008tw}.
For instance, $\nu_{R}\nu_{R}$, $e_{L}\nu_{R}$, and $\nu_{L}^{\dagger}\nu_{L}^{\dagger}$
are Lorentz invariant  under the implied contraction of spinor indices:
$\nu_{R}\nu_{R}\equiv\nu_{R}^{\alpha}\nu_{R\alpha}$, $e_{L}\nu_{R}\equiv e_{L}^{\alpha}\nu_{R\alpha}$,
and $\nu_{L}^{\dagger}\nu_{L}^{\dagger}\equiv\nu_{L\dot{\alpha}}^{\dagger}\nu_{L}^{\dagger\dot{\alpha}}$,
where $\alpha$ and $\dot{\alpha}$ are the so-called {\it dotted
and undotted} spinor indices~\cite{Dreiner:2008tw}. 

In this paper, we mainly use Weyl-van der Waerden spinors for their
simplicity. Since Dirac spinors are more extensively used in the literature,
we also provide here the conversion between the two notations for
the reader's convenience. We denote the Dirac spinor of a neutrino
by $\Psi$, which can be decomposed into two two-component spinors:
\begin{equation}
\Psi=\left(\begin{array}{c}
P_{2}\Psi_{L}\\
P_{2}\Psi_{R}
\end{array}\right)=\left(\begin{array}{c}
\left(\nu_{L}\right)_{\alpha}\\[2mm]
\left(\nu_{R}^{\dagger}\right)^{\dot{\alpha}}
\end{array}\right),\ \ \Psi^{C}\equiv\left(\begin{array}{c}
\left(\nu_{R}\right)_{\alpha}\\[2mm]
\left(\nu_{L}^{\dagger}\right)^{\dot{\alpha}}
\end{array}\right),\label{eq:-44}
\end{equation}
where $\Psi_{L,R}\equiv P_{L,R}\Psi$ with  $P_{L,R}=(1\mp\gamma^{5})/2$,
$P_{2}$ denotes the projector that selects two nonzero components
of the four-component spinors $\Psi_{L}$ or $\Psi_{R}$, and $\Psi^{C}$
is the charge conjugate of $\Psi$. With the decomposition in Eq.~\eqref{eq:-44},
the Yukawa term and its hermitian conjugate in Eq.~\eqref{eq:L} can
be written in terms of Dirac spinors as follows: 
\begin{align}
\nu_{R}\nu_{R}\phi & =\overline{\Psi}P_{L}\Psi^{C}\phi=\overline{\Psi_{R}}\Psi_{L}^{C}\phi\thinspace,\label{eq:-45}\\
\nu_{R}^{\dagger}\nu_{R}^{\dagger}\phi & =\overline{\Psi^{C}}P_{R}\psi\phi=\overline{\Psi_{L}^{C}}\Psi_{R}\phi\thinspace,\label{eq:-46}
\end{align}
where $\Psi_{L,R}^{C}=P_{L,R}\Psi^{C}$. For other terms, the conversion
is similar. For instance, the Dirac mass terms can be written as $\nu_{L}\nu_{R}=\nu_{R}\nu_{L}=\overline{\Psi}P_{L}\Psi$,
 $\nu_{L}^{\dagger}\nu_{R}^{\dagger}=\nu_{R}^{\dagger}\nu_{L}^{\dagger}=\overline{\Psi}P_{R}\Psi$,
and $\nu_{L}\nu_{R}+\nu_{L}^{\dagger}\nu_{R}^{\dagger}=\overline{\Psi}\Psi$. 

Using Weyl-van der Waerden spinors, it is also straightforward to
construct the Majorana spinor of $\nu_{R}$:
\begin{equation}
\Psi_{{\cal N}}=\left(\begin{array}{c}
\left(\nu_{R}\right)_{\alpha}\\[2mm]
\left(\nu_{R}^{\dagger}\right)^{\dot{\alpha}}
\end{array}\right),\ \Psi_{{\cal N}}^{C}=\Psi_{{\cal N}}\thinspace.\label{eq:-47}
\end{equation}
Then Eqs.~\eqref{eq:-45} and \eqref{eq:-46} can be written as 
\begin{equation}
\nu_{R}\nu_{R}\phi+\nu_{R}^{\dagger}\nu_{R}^{\dagger}\phi=\overline{\Psi_{{\cal N}}}P_{L}\Psi_{{\cal N}}\phi+\overline{\Psi_{{\cal N}}}P_{R}\Psi_{{\cal N}}\phi=\overline{\Psi_{{\cal N}}}\Psi_{{\cal N}}\phi\thinspace.\label{eq:-48}
\end{equation}
Due to $\Psi_{{\cal N}}^{C}=\Psi_{{\cal N}}$, the quantization of
$\Psi_{{\cal N}}$ gives rise to particles (denoted by ${\cal N}$)
that are exactly their own antiparticles.  We emphasize here that
our notation distinguishes particles (${\cal N}$) from fields ($\nu_{R}$
or $\Psi$). An important difference in practical use is that the
latter allow one to take hermitian conjugates ($\nu_{R}^{\dagger}$
or $\overline{\Psi}$), while such conjugate notations should be absent
for ${\cal N}$. Therefore, the process of two right-handed neutrinos
annihilating to $\phi$ particles is expressed as ${\cal N}+{\cal N}\to\phi+\phi$
rather than $\nu_{R}+\nu_{R}\to\phi+\phi$ because the latter could
be confusing in the presence of the notation $\nu_{R}^{\dagger}$.

At low-energy scales after electroweak symmetry breaking, ${\cal L}_{{\rm seesaw}}$
generates neutrino masses via the well-known seesaw relation
\begin{equation}
m_{\nu}=\frac{m_{D}^{2}}{m_{R}}\thinspace,\ m_{D}=\frac{y_{\nu}v}{\sqrt{2}}\thinspace,\label{eq:-1}
\end{equation}
where $v\approx246$ GeV is the electroweak vacuum expectation value
(VEV), $m_{D}$ is the Dirac mass of neutrinos generated by the Higgs
mechanism, and $m_{\nu}={\cal O}(0.1)$ eV is the neutrino mass responsible
for neutrino oscillations. 

Throughout this work, we assume that the $\nu_{R}$-SM coupling $y_{\nu}$
is determined by Eq.~\eqref{eq:-1}, i.e., $y_{\nu}=\sqrt{2m_{\nu}m_{R}}/v$.

\subsection{The Boltzmann equation}

In the expanding universe, the number density of $\phi$, denoted
by $n_{\phi},$ is governed by the Boltzmann equation:
\begin{equation}
\frac{dn_{\phi}}{dt}+3Hn_{\phi}=C_{\text{prod}}-C_{\text{depl}}\thinspace,\label{eq:-10}
\end{equation}
where $H$ is the Hubble parameter, and $C_{\text{prod}}$ and $C_{\text{depl}}$
denote the collision terms responsible for the production and depletion
of $\phi$, respectively. Since the left-hand side can also be written
as $a^{-3}d(n_{\phi}a^{3})/dt$, $C_{\text{prod}}$ and $C_{\text{depl}}$
can be physically interpreted as  the number of $\phi$ being produced
and depleted per unit time per comoving volume. 

For convenience, we introduce $g_{H}\equiv\sqrt{8\pi^{3}g_{\star}/90}$
with $g_{\star}$ the effective number of relativistic degrees of
freedom in the thermal bath, such that
\begin{equation}
H=g_{H}\frac{T^{2}}{m_{{\rm pl}}}\thinspace,\label{eq:-11}
\end{equation}
where $T$ is the temperature of the SM thermal bath and $m_{{\rm pl}}=1.22\times10^{19}$
GeV is the Planck mass. 

If $\phi$ reaches thermal equilibrium, its number density is given
by
\begin{equation}
n_{\phi}^{{\rm eq}}=\int\frac{1}{e^{E/T}-1}\frac{d^{3}p}{(2\pi)^{3}}=\begin{cases}
\zeta(3)\frac{T^{3}}{\pi^{2}} & (T\gg m_{\phi})\\[2mm]
\left(\frac{m_{\phi}T}{2\pi}\right)^{3/2}e^{-m_{\phi}/T} & (T\ll m_{\phi})
\end{cases}\thinspace,\label{eq:n-eq}
\end{equation}
where $\zeta$ is the Riemann zeta function and $\zeta(3)\approx1.202$. 

The DM relic abundance $\Omega_{\phi}h^{2}$ is related to $n_{\phi}$
via
\begin{equation}
\Omega_{\phi}h^{2}=0.12\times\left.\frac{m_{\phi}n_{\phi}}{\rho_{{\rm DM},0}}\right|_{{\rm today}}\thinspace,\label{eq:-14}
\end{equation}
where $\rho_{{\rm DM},0}=9.74\times10^{-12}\ \text{eV}^{4}$ is the
DM energy density of the universe today. 

\section{Viability as a DM candidate\label{sec:Viability}}

\subsection{Lifetime of $\phi$\label{subsec:Lifetime}}

\begin{figure}

\centering

\includegraphics[width=0.95\textwidth]{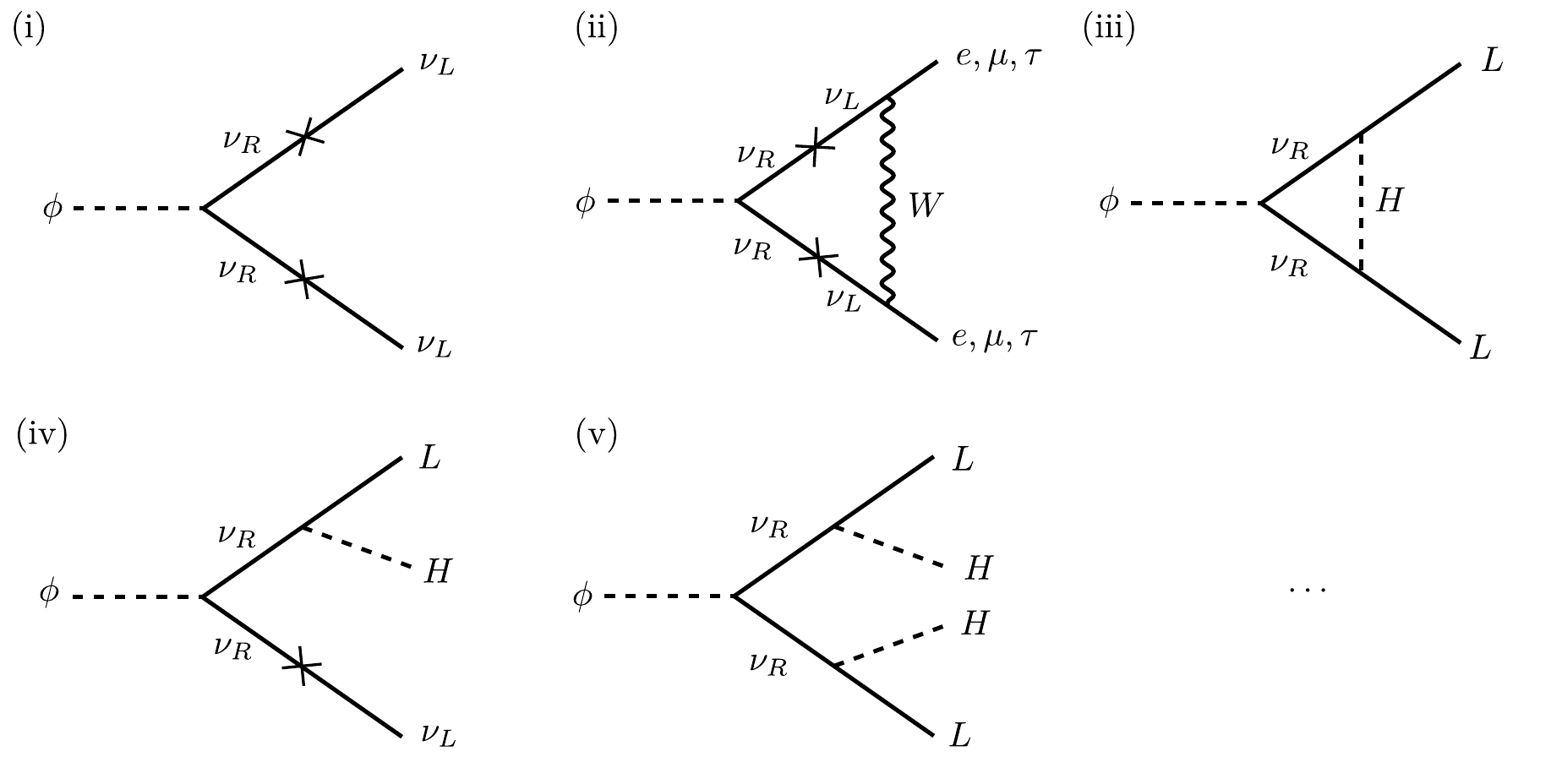}

\caption{Feynman diagrams for $\phi$ decay. The mass insertion ``$\times$''
on the neutrino lines denote the Dirac mass term connecting $\nu_{L}$
and $\nu_{R}$. \label{fig:Feynman}}

\end{figure}

The $\nu_{R}$-philic scalar $\phi$, indirectly coupled to the SM
via $y\nu_{R}\nu_{R}\phi$ and $y_{\nu}\tilde{H}^{\dagger}L\nu_{R}$,
is not absolutely stable. To qualify as DM, its lifetime $\tau_{\phi}$
must be at least longer than the age of the universe, $\tau_{{\rm univ}}$.

If $m_{\phi}>2m_{R}$, it would decay into two right-handed neutrinos,
rendering the lifetime of $\phi$ in general much shorter than $\tau_{{\rm univ}}$,
unless $y$ is extremely small (e.g. below $10^{-21}$ for GeV $m_{\phi}$).
 We therefore only consider $m_{\phi}$ below the seesaw scale, making
$\phi\to2{\cal N}$ kinematically forbidden. As a result, $\phi$
becomes generally much more long-lived.

In this scenario, multiple decay channels can be responsible for the
lifetime of $\phi$. In Fig.~\ref{fig:Feynman} we enumerate the
most relevant ones and discuss them as follows.

Diagram (i) in Fig.~\ref{fig:Feynman} is the simplest and, within
certain ranges of $m_{\phi}$ and $m_{R}$, also the dominant process
of $\phi$ decay. Assuming $m_{R}\gg m_{\phi}\gg m_{\nu}$, the decay
width is approximately given by
\begin{equation}
\Gamma_{\phi\to2\nu_{L}}\approx N_{f}\frac{y^{2}m_{\phi}}{16\pi}\left(\frac{m_{D}}{m_{R}}\right)^{4},\label{eq:-2}
\end{equation}
where $N_{f}=3$ is the number of lepton flavors. 

Diagrams (ii) and (iii) are loop-level processes. They are of  the
same order of magnitude. In fact, since the electroweak gauge symmetry
is spontaneously broken, the two diagrams should be taken into account
together, otherwise the results would be gauge dependent---see Ref.~\cite{Xu:2020qek}
for the cancellation of the gauge dependence between the two diagrams.
In the heavy $\nu_{R}$ limit,   the two diagrams give rise to the
following effective couplings to charged leptons~\cite{Xu:2020qek}\footnote{Note that here $y_{\phi\ell\ell}\propto m_{\ell}m_{\nu}$ is a consequence
of chirality flipping. If $\phi$ is replaced with a vector boson,
then the loop-induced coupling would be proportional to $m_{D}^{2}$
instead of $m_{\ell}m_{\nu}$~\cite{Chauhan:2020mgv,Chauhan:2022iuh}. }:
\begin{equation}
y_{\phi\ell\ell}=\frac{3y}{16\sqrt{2}\pi^{2}}G_{F}m_{\ell}m_{\nu}\thinspace,\label{eq:-3}
\end{equation}
where $\ell=e$, $\mu$ or $\tau$, with the mass denoted by $m_{\ell}$.
In addition to the above loop diagrams, $\phi$ could also be indirectly
coupled to SM fermions via a neutrino loop and a $Z$ mediator. It
turns out that such a diagram is proportional to $y-y^{*}$ which
vanishes for real $y$---see Eqs.~(C8)-(C9) in Ref.~\cite{Xu:2020qek}.
So the $Z$-mediated loop diagram can be neglected in this work. 

Using Eq.~\eqref{eq:-3}, we obtain the decay width of $\phi$ to
two charged leptons for $m_{\phi}\gg m_{{\rm \ell}}$:
\begin{equation}
\Gamma_{\phi\to2\ell}\approx\frac{9y^{2}m_{\phi}}{2^{12}\pi^{5}}\left(G_{F}m_{\ell}m_{\nu}\right)^{2}.\label{eq:-4}
\end{equation}

It is interesting to note that sometimes the loop-level processes
may dominate the decay. This can be seen by comparing Eq.~\eqref{eq:-4}
to Eq.~\eqref{eq:-2}:
\begin{equation}
\frac{\Gamma_{\phi\to2\ell}}{\Gamma_{\phi\to2\nu_{L}}}\approx\left(\frac{m_{R}}{\Lambda_{\ell}}\right)^{2},\ \text{ with }\Lambda_{\ell}\equiv\frac{16\sqrt{2}\pi^{2}}{3G_{F}m_{\ell}}=\begin{cases}
1.5\times10^{10}\ \text{GeV} & (\ell=e)\\
7.4\times10^{7}\ \text{GeV} & (\ell=\mu)\\
4.4\times10^{6}\ \text{GeV} & (\ell=\tau)
\end{cases}\thinspace,\label{eq:-5}
\end{equation}
which implies that $\Gamma_{\phi\to2\ell}$ can dominate over $\Gamma_{\phi\to2\nu_{L}}$
if $m_{R}$ is above the scale $\Lambda_{\ell}$ and if $\phi\to2\ell$
is kinematically allowed ($m_{\phi}>2m_{\ell}$). 

The reason for the loop-level processes being dominant at large $m_{R}$
is that, compared to diagram (i), these  loop diagrams are less suppressed
by $1/m_{R}$. Although they all contain two heavy $\nu_{R}$ propagators
which lead to a suppression of $(1/m_{R})^{2}$, the latter also contains
a loop integral in which the loop momentum may effectively  run up
to the scale of $m_{R}$.  Consequently, it reduces the power of
$1/m_{R}$, as implied by Eq.~\eqref{eq:-3}, where $m_{\nu}=m_{D}^{2}/m_{R}$
can be understood as a suppression of only $1/m_{R}$. 

Next to the above two-body decay processes, there are three- and four-body
decay processes, which could  also dominate the decay under certain
conditions. Diagrams (iv) and (v) in Fig.~\ref{fig:Feynman} can
be viewed as variants of diagram (i) with one or two mass insertions
being replaced by Higgs legs. If the decay is sufficiently energetic
($m_{\phi}\gg v$) such that the electroweak VEV $v$ can be regarded
as a small quantity, it is likely that part of the energy tends to
be released via extra Higgs emission to overcome the suppression caused
by the VEV (or equivalently, the mass insertion $m_{D}\propto v$).
This feature has previously been noticed and discussed in Ref.~\cite{Dudas:2014bca}. 

More specifically, let us compare the squared amplitudes of diagrams
(iv) and (i), $|{\cal M}_{(\text{iv})}|^{2}/|{\cal M}_{(\text{i})}|^{2}\sim(y_{\nu}/m_{D})^{2}\sim1/v^{2}$.
This difference should be compensated by an extra phase space integral
which typically contributes a factor of $m_{\phi}^{2}/(16\pi^{2})$.
Hence by simple estimation, we obtain that the decay width of (iv)
dominates over (i) when $m_{\phi}^{2}/v^{2}\gg16\pi^{2}$. However,
 as we will show later, if we demand that the coupling $y$ is responsible
for the production of $\phi$ in the early universe, there is no viable
parameter space for $m_{\phi}$ above the electroweak scale. Hence
diagrams (iv) and (v) are irrelevant to our analysis.

\subsection{Thermal production of $\phi$ in the early universe\label{subsec:Thermal-production}}

\subsubsection{Production via $2{\cal N}\to2\phi$}

Now let us consider the thermal production of $\phi$ in the early
universe. Since $\phi$ is only coupled to right-handed neutrinos
which were  abundant in the early universe at temperatures above or
around $m_{R}$, the simplest process that can efficiently produce
$\phi$ (with $m_{\phi}<m_{R}$) is $2{\cal N}\to2\phi$.  So let
us first discuss the thermal production of $\phi$ via this process. 

It is important to note that the seesaw Yukawa coupling $y_{\nu}$
is in general strong enough to keep $\nu_{R}$ in thermal equilibrium~\cite{Li:2022bpp},
provided that $y_{\nu}$ is related to neutrino masses via the seesaw
relation, $y_{\nu}=\sqrt{2m_{\nu}m_{R}}/v$. Therefore, in our calculation
of the thermal production of $\phi$, the number density of $\nu_{R}$
takes the equilibrium value\footnote{Strictly speaking, since the reaction rate of $\nu_{R}\leftrightarrow LH$
is only about ${\cal O}(10)$ times the Hubble expansion rate at $T\sim m_{R}$,
$\nu_{R}$ is slightly out of equilibrium when it becomes nonrelativistic.
The small deviation from equilibrium is crucial to leptogenesis~\cite{Buchmuller:2004nz}
but negligible in our analysis. }.

The production rate of $\phi$, denoted by $C_{{\rm prod}}$ in Eq.~\eqref{eq:-10},
  is a dimension-four quantity. At sufficiently high temperatures
when all particle masses are negligible, the  contribution of  a
two-to-two process to $C_{{\rm prod}}$ should be roughly proportional
to $T^{4}$ according to dimensional analysis. Indeed, for the $2{\cal N}\to2\phi$
process,  we find that its contribution to $C_{{\rm prod}}$ in the
high-$T$ limit is given by
\begin{equation}
C_{2{\cal N}\to2\phi}\approx\frac{y^{4}N_{f}}{128\pi^{5}}T^{4}\times4\ln\left(\frac{2T}{m_{R}}\right)\ \ \text{for}\ T\gg m_{R}\thinspace.\label{eq:-6}
\end{equation}

At $T\ll m_{R}$, $\nu_{R}$ becomes nonrelativistic with the number
density suppressed by $e^{-m_{R}/T}$. Consequently, the production
rate is suppressed by $e^{-2m_{R}/T}$. Using the non-relativistic
approximation, we obtain
\begin{equation}
C_{2{\cal N}\to2\phi}\sim\frac{y^{4}N_{f}}{128\pi^{5}}T^{4}\times9\pi e^{-2m_{R}/T}\ \ \text{for}\ T\ll m_{R}\thinspace.\label{eq:-7}
\end{equation}

For general $T$, $C_{2{\cal N}\to2\phi}$ cannot be analytically
computed but the numerical calculation of $C_{2{\cal N}\to2\phi}$
is straightforward\footnote{We use the Monte-Carlo code available at \url{https://github.com/xunjiexu/Thermal_Monte_Carlo}~\cite{Luo:2020fdt}.}.
In practice, we find that the following expression, obtained by analytically
incorporating the high- and low-$T$ limits in Eqs.~\eqref{eq:-6}
and \eqref{eq:-7}, agrees well with the numerical result (see Appendix~\ref{sec:M-and-C},
Fig.~\ref{fig:monte}):
\begin{equation}
C_{2{\cal N}\to2\phi}\approx\frac{y^{4}N_{f}}{128\pi^{5}}T^{4}\times4\ln\left(\frac{2T}{m_{R}}+\frac{e^{9\pi/4}}{1+(2T/m_{R})^{2}}\right)e^{-2m_{R}/T}\thinspace.\label{eq:-8}
\end{equation}

If $y$ is sufficiently small, one can use the freeze-in formula to
evaluate the number density of $\phi$ produced via this process~\cite{Li:2022bpp}:
\begin{equation}
n_{\phi}^{(1)}=T^{3}\frac{m_{{\rm pl}}}{g_{H{\rm f.i.}}}\frac{g_{\star}}{g_{\star{\rm f.i.}}}\int_{T}^{\infty}\frac{C_{2{\cal N}\to2\phi}(T')}{T'^{6}}dT'\thinspace,\label{eq:-9}
\end{equation}
where $\text{\ensuremath{g_{H{\rm f.i.}}} and }$$g_{\star{\rm f.i.}}$
denote the freeze-in values of $g_{H}$ and $g_{\star}$, respectively.
For $m_{R}$ well above the electroweak scale, we take $g_{\star{\rm f.i.}}=106.75$
and $g_{H{\rm f.i.}}=17.15$. 

Substituting Eq.~\eqref{eq:-8} into Eq.~\eqref{eq:-9} and computing
the integral, we obtain
\begin{equation}
\frac{n_{\phi}^{(1)}}{T^{3}}\approx0.046\left(\frac{y}{10^{-2}}\right)^{4}\left(\frac{10^{8}\ \text{GeV}}{m_{R}}\right)\frac{g_{\star}}{106.75}\thinspace.\label{eq:-12}
\end{equation}
The superscript ``$^{(1)}$'' is introduced to distinguish the $2{\cal N}\to2\phi$
contribution from other contributions considered later. 

The corresponding contribution to the relic abundance of DM according
to Eq.~\eqref{eq:-14} is 
\begin{equation}
\Omega_{\phi}^{(1)}h^{2}=0.12\times\left(\frac{y}{10^{-4}}\right)^{4}\frac{m_{\phi}/m_{R}}{4.6\times10^{-7}}\thinspace.\label{eq:-15}
\end{equation}

The validity of Eq.~\eqref{eq:-12} {[}and hence Eq.~\eqref{eq:-15}{]}
relies on the assumption that $\phi$ does not thermalize, which requires
$n_{\phi}^{(1)}\ll\zeta(3)T^{3}/\pi^{2}$, corresponding to $y\ll y_{{\rm eq}}$
with 
\begin{equation}
y_{{\rm eq}}\approx0.013\left(\frac{m_{R}}{10^{8}\ \text{GeV}}\right)^{1/4}\thinspace.\label{eq:-13}
\end{equation}
For $y\gg y_{{\rm eq}}$, the process $2{\cal N}\leftrightarrow2\phi$
would maintain $\phi$ in thermal equilibrium with the thermal bath
until $T$ drops well below $m_{R}$. As we will discuss  later, once
$\phi$ enters equilibrium, no mechanism can significantly reduce
the number of $\phi$ particles in the  comoving volume. So $n_{\phi}a^{3}$
remains constant in the subsequent evolution, while $n_{\phi}/T^{3}$
is subject to entropy dilution caused by heavy SM species annihilating
or decaying into lighter species. 

If $n_{\phi}$ has reached its equilibrium value before the process
$2{\cal N}\leftrightarrow2\phi$ terminates, and if this happens above
the electroweak scale,  then the number density of $\phi$ today
is given by 
\begin{equation}
n_{\phi,0}=T_{0}^{3}\frac{\zeta(3)}{\pi^{2}}\frac{g_{{\rm \star}0}}{106.75}\thinspace,\label{eq:-16}
\end{equation}
where $T_{0}\approx2.7\ \text{K}$ is the CMB temperature and $g_{{\rm \star}0}=2+7/8\times6\times4/11\approx3.91$
is the effective value of $g_{\star}$ after taking neutrino decoupling
and $e^{+}e^{-}$ annihilation into account. Substituting Eq.~\eqref{eq:-16}
into Eq.~\eqref{eq:-14}, we obtain that $\Omega_{\phi}h^{2}$ would
exceed $0.12$ if 
\begin{equation}
m_{\phi}>173.4\ {\rm eV}\thinspace.\label{eq:th-bound}
\end{equation}
Therefore, Eq.~\eqref{eq:th-bound} should be be considered as the
lower bound of $m_{\phi}$ for $\phi$ to serve as a DM candidate,
provided that $\phi$ has decoupled from the thermal bath before $T$
drops below the electroweak scale.

\subsubsection{Production via $LH\to{\cal N}\phi$}

The process $LH\to{\cal N}\phi$ can be important if the seesaw Yukawa
coupling $y_{\nu}$ is comparable or larger than $y$. Since the Feynman
diagram of $LH\to{\cal N}\phi$ contains two vertices proportional
to $y$ and $y_{\nu}$, the production rate via this process is proportional
to $y^{2}y_{\nu}^{2}$. Performing a similar calculation, we find
that the corresponding collision term can be approximately given by
\[
C_{LH\to{\cal N}\phi}\approx\frac{y^{2}y_{\nu}^{2}N_{f}}{64\pi^{5}}\begin{cases}
T^{4}\thinspace & \ \ \text{for}\ T\gg m_{R}\thinspace\\
3m_{R}^{2}T^{2}K_{2}\left(\frac{m_{R}}{T}\right) & \ \ \text{for}\ T\ll m_{R}\thinspace
\end{cases}.
\]
Here,  for the sake of simplicity in notation, we have combined the
contributions of $LH\to{\cal N}\phi$ and $\overline{L}\overline{H}\to{\cal N}\phi$
into a single collision term, $C_{LH\to{\cal N}\phi}$. 

Similar to the procedure of deriving Eq.~\eqref{eq:-8}, the high-$T$
and low-$T$ limits can be combined to obtain an analytical expression
which  is sufficiently accurate for general $T$:
\begin{equation}
C_{LH\to{\cal N}\phi}\approx\frac{y^{2}y_{\nu}^{2}N_{f}}{64\pi^{5}}3m_{R}^{2}T^{2}K_{2}\left(\frac{m_{R}}{T}\right)\frac{m_{R}^{2}+T^{2}}{m_{R}^{2}+6T^{2}}\thinspace.\label{eq:-17}
\end{equation}
Then by integrating over $T$, we obtain the $LH\to{\cal N}\phi$
contribution to the final number density of $\phi$:
\begin{equation}
\frac{n_{\phi}^{(2)}}{T^{3}}\approx0.063\left(\frac{yy_{\nu}}{10^{-4}}\right)^{2}\left(\frac{10^{8}\ \text{GeV}}{m_{R}}\right)\frac{g_{\star}}{106.75}\thinspace.\label{eq:-18}
\end{equation}
Obviously $n_{\phi}^{(2)}$ would dominate over $n_{\phi}^{(1)}$
in Eq.~\eqref{eq:-12} if $y_{\nu}/y\gtrsim0.85$. 

The corresponding contribution to $\Omega_{\phi}h^{2}$, according
to Eq.~\eqref{eq:-14}, is 
\begin{equation}
\Omega_{\phi}^{(2)}h^{2}=0.12\times\left(\frac{yy_{\nu}}{10^{-8}}\right)^{2}\frac{m_{\phi}/m_{R}}{3.3\times10^{-7}}\thinspace.\label{eq:-19}
\end{equation}
In our calculation of the DM relic abundance, we include both contributions
from $2{\cal N}\to2\phi$ and $LH\to{\cal N}\phi$, i.e. $\Omega_{\phi}h^{2}=\Omega_{\phi}^{(1)}h^{2}+\Omega_{\phi}^{(2)}h^{2}$.

\subsubsection{Other processes }

In addition to $LH\to{\cal N}\phi$ and $2{\cal N}\to2\phi$, there
are other processes that could produce $\phi$ but their contributions
are negligible, as we shall discuss below. 

Inverse decay processes, such as $2\nu_{L}\to\phi$ or $2\ell\to\phi$,
generally have production rates suppressed by the decay widths of
$\phi$, which have to be small for $\phi$ to quality as DM. Taking
$2\nu_{L}\to\phi$ as an example, the production rate reads~\cite{EscuderoAbenza:2020cmq}:
\begin{equation}
C_{2\nu_{L}\to\phi}\approx\Gamma_{\phi\to2\nu_{L}}n_{\phi}^{{\rm eq}}\frac{K_{1}(x)}{K_{2}(x)}\thinspace,\label{eq:-20}
\end{equation}
where $x\equiv m_{\phi}/T$ and $K_{1,2}$ are modified Bessel functions.
The ratio $K_{1}(x)/K_{2}(x)$ approaches $x/2$ and $1$ in the $x\to0$
and $x\to\infty$ limits, respectively. To satisfy $\tau_{\phi}>\tau_{{\rm univ}}$,
 we require $\Gamma_{\phi\to2\nu_{L}}<\tau_{{\rm univ}}^{-1}\sim H_{0}$
where $H_{0}$ is the Hubble parameter today. Due to $H_{0}\ll H$
in the early universe and $K_{1}(x)/K_{2}(x)<1$, we always have $C_{2\nu_{L}\to\phi}\ll Hn_{\phi}^{{\rm eq}}$,
which implies that the production rate is negligibly small. Alternatively,
one could see this by performing an integration similar to Eq.~\eqref{eq:-9}
for $C_{2\nu_{L}\to\phi}$. The resulting yield of $\phi$ is much
lower than that from Eq.~\eqref{eq:-9}. The same conclusion also
holds for any other inverse decay processes, as long as $\tau_{\phi}>\tau_{{\rm univ}}$
is imposed.

Two-to-two scattering processes such as $2\nu_{L}\to2\phi$ or $2\ell\to2\phi$
should have lower production rates than the aforementioned inverse
decay processes because adding extra $\phi$ legs to the Feynman diagrams
of the latter generates exactly the diagrams of these two-to-two processes.
 This is based on the general argument of power counting of $y$,
which could be invalid in the presence of helicity suppression. Nevertheless,
by analyzing the effective operators responsible for this process,
one can see that these processes are indeed negligible for the cosmological
evolution of $\phi$. 

In fact, at high temperatures above $m_{R}$, no two-to-two scattering
processes can be more efficient than $LH\leftrightarrow{\cal N}\phi$
and $2{\cal N}\leftrightarrow2\phi$ which involve the minimal number
of $y$ or $y_{\nu}$ vertices. At low temperatures below $m_{R}$,
one can integrate out $\nu_{R}$ and consider only effective operators.
Taking $2\nu_{L}\leftrightarrow2\phi$ for example, after integrating
out $\nu_{R}$, we obtain the effective operator: 
\begin{equation}
{\cal L}_{{\rm eff}}=\frac{1}{\Lambda}\nu_{L}\nu_{L}\phi\phi\ \ \text{with}\ \ \Lambda^{-1}\sim y^{2}\frac{m_{D}^{2}}{m_{R}^{3}}\thinspace.\label{eq:-21}
\end{equation}
In analogy with the standard neutrino decoupling at $T\sim(G_{F}^{2}m_{\text{pl}})^{-1/3}$~\cite{Luo:2020sho},
we see that the effective interaction in Eq.~\eqref{eq:-21} implies
a decoupling temperature at 
\begin{equation}
T\sim(\Lambda^{-2}m_{\text{pl}})^{-1}.\label{eq:-22}
\end{equation}
 As we have checked, as long as $y$ respects the unitarity bound
and $m_{D}$ is determined by the seesaw relation,   the decoupling
temperature in Eq.~\eqref{eq:-22} would always be well above $m_{R}$.
This implies that $2\nu_{L}\leftrightarrow2\phi$ at $T<m_{R}$ can
never reach equilibrium. Note that here we assume both $\nu_{L}$
and $\phi$ in $2\nu_{L}\leftrightarrow2\phi$ are relativistic. Non-relativistic
$\phi$ would lead to a more suppressed reaction rate. Therefore,
we conclude that $2\nu_{L}\leftrightarrow2\phi$ is negligible in
determining the relic abundance of $\phi$. In particular, $n_{\phi}a^{3}$
 at $T<m_{R}$ cannot be significantly increased by $2\nu_{L}\to2\phi$,
nor decreased by the backreaction, $2\phi\to2\nu_{L}$. Consequently,
the freeze-out mechanism cannot be accommodated here. For other two-to-two
processes, we arrive at the same conclusion after a similar analysis.

\subsection{Parameter space}

\begin{figure}
\centering

\phantom{xxxxxxxxxx} \includegraphics[width=0.8\textwidth]{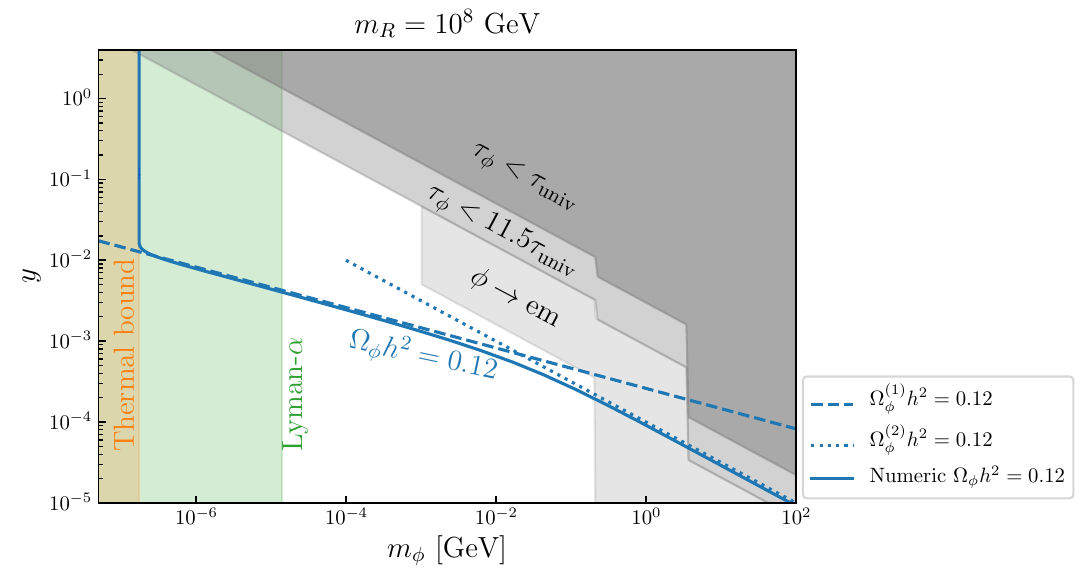}

\includegraphics[width=0.99\textwidth]{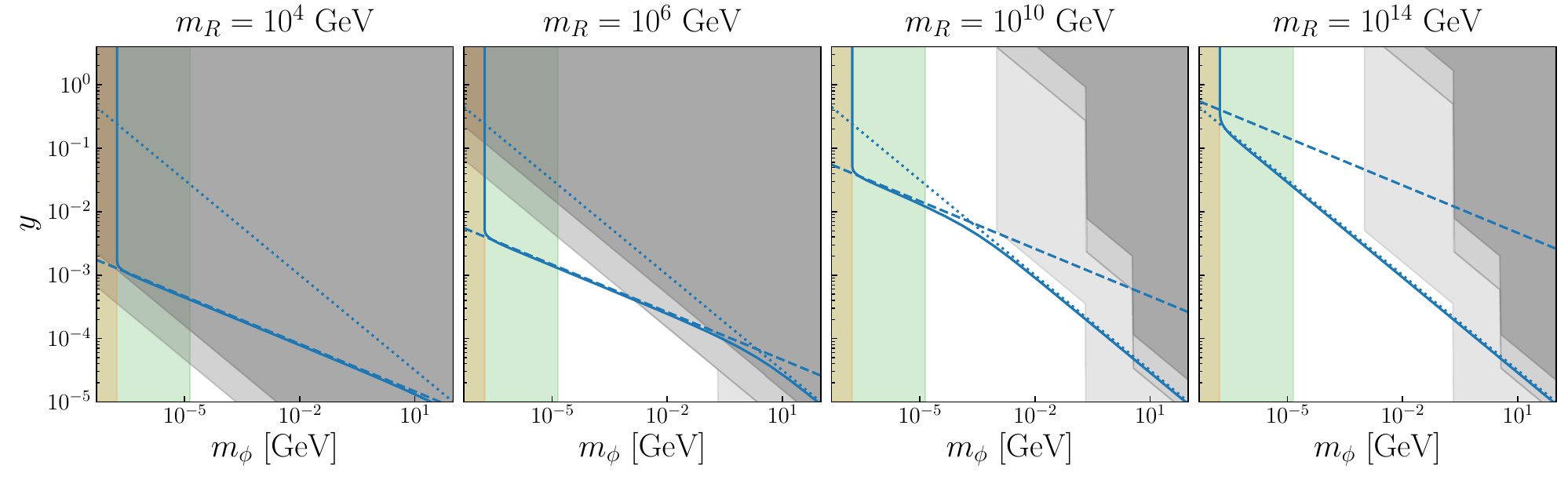}

\caption{Upper panel: the parameter space of the $\nu_{R}$-philic scalar DM,
assuming $m_{R}=10^{8}$ GeV. The blue solid curve leads to the observed
DM relic abundance, $\Omega_{\phi}h^{2}=0.12$, obtained by numerically
solving the Boltzmann equation. The dashed and dotted lines take into
account only the contributions of $2{\cal N}\to2\phi$ and $LH\to{\cal N}\phi$,
obtained using Eqs.~\eqref{eq:-15} and \eqref{eq:-19}, respectively.
 The gray shaded regions correspond to $\tau_{\phi}<\tau_{{\rm univ}}$
(darkest gray region), $\tau_{\phi}<11.5\ \tau_{{\rm univ}}$ (lighter
gray region), and $\tau_{\phi\to{\rm em}}<10^{25}\ \text{s}$ (lightest
gray region). The thermal bound (orange) applies to DM candidates
that are produced via thermal processes above the electroweak scale---see
Eq.~\eqref{eq:th-bound}. The Lyman-$\alpha$ bound (green) is derived
by requiring that the free-streaming length is below a certain limit.
 Lower panels: similar to the upper panel but with different values
of $m_{R}$. The color coding is the same as the upper panel. \label{fig:para-space}}
\end{figure}

In our model, there are only three relevant free parameters: the coupling
$y$ and the two masses $m_{\phi}$ and $m_{R}$. The seesaw Yukawa
coupling $y_{\nu}$ is essentially determined by $m_{R}$ via the
seesaw relation in Eq.~\eqref{eq:-1}. Therefore, a convenient way
to explore the parameter space is to fix $m_{R}$ (consequently, $y_{\nu}=\sqrt{2m_{\nu}m_{R}}/v$
is also fixed) and vary $m_{\phi}$ and $y$.

We are mainly concerned with three constraints: 

(i) large $y$ and $m_{\phi}$ may render the lifetime of $\phi$
too short; 

(ii) large $y$ may bring $\phi$ into thermal equilibrium and, in
the absence of freeze-out suppression,   may potentially lead to
an overproduction of $\Omega_{\phi}h^{2}$;

(iii)  small $m_{\phi}$ leads to a long free-streaming length at
the epoch of structure formation and thus can be constrained by Lyman-$\alpha$
observations.

For constraint (i), we include the dominant decay widths calculated
in Sec.~\ref{subsec:Lifetime} and compare the obtained lifetime
with the age of the universe, $\tau_{{\rm univ}}\approx4.35\times10^{17}\ \text{s}$
($13.79\times10^{9}$ yr).  This is presented in Fig.~\ref{fig:para-space},
where the darkest gray region corresponds to $\tau_{\phi}<\tau_{{\rm univ}}$.
One should note, however, that this is only a conservative theoretical
bound. When being confronted with observational data, $\tau_{\phi}$
actually needs to be much larger than $\tau_{{\rm univ}}$. The specific
bounds depend on whether $\phi$ decays to invisible radiations (e.g.~$\nu_{L}$)
or electromagnetic channels (e.g.~$e^{\pm}$, $\mu^{\pm}$, $\tau^{\pm}$,
$\gamma$). For the former, the latest bound is $\tau_{\phi}^{-1}<6.3\times10^{-3}\ \text{Gyr}^{-1}$~\cite{Poulin:2016nat},
corresponding to $\tau_{\phi}>11.5\ \tau_{{\rm univ}}$, which is
shown in Fig.~\ref{fig:para-space} as the lighter gray region. For
the latter, the bounds depend on the ionization efficiency of the
decay~\cite{Zhang:2007zzh}. For instance, $\nu_{L}$ from $\phi$
decay or from the subsequent decay of $\mu^{\pm}$ and $\tau^{\pm}$
cannot contribute to the ionization energy. If the decay energy is
fully converted to the ionization energy, then the decay width  should
be below $10^{-25}\ \text{s}^{-1}$~\cite{Poulin:2016anj}. In our
model, the total decay width of $\phi$ consists of contributions
from $\phi\to2\nu_{L}$ and $\phi\to2\ell$. Only the latter causes
effective energy injection into the electromagnetic sector. Hence
the $10^{-25}\ \text{s}^{-1}$ bound should be applied only to $\Gamma_{\phi\to2\ell}$.
We  present this bound in Fig.~\ref{fig:para-space} as the lightest
gray region.

For constraint (ii), since the dominant processes for $\phi$ production
 is efficient only at temperatures above the electroweak scale while
the decay is significant only at a time scale much larger than $\tau_{{\rm univ}}$,
we assume that $n_{\phi}a^{3}$ remains constant during the period
from $T\sim{\cal O}(10^{2})$ GeV to today. Note that the maximum
of $n_{\phi}a^{3}$ is $n_{\phi}^{{\rm eq}}a^{3}=\frac{\zeta(3)}{\pi^{2}}T^{3}a^{3}$
according to Eq.~\eqref{eq:n-eq}. The corresponding maximal value
of $n_{\phi}$ today then implies a lower bound on $m_{\phi}$ according
to Eq.~\eqref{eq:-14}. This bound is given by Eq.~\eqref{eq:th-bound}
and indicated as the thermal bound in Fig.~\ref{fig:para-space}.

For constraint (iii), the bound on $m_{\phi}$ can be obtained by
computing the free-streaming length and comparing it with that of
other DM scenarios with known Lyman-$\alpha$ bounds.  For a generic
particle with mass $m$, the free-streaming length is given by~\cite{Coy:2021ann,Wang:2023csv}
\begin{equation}
\lambda_{{\rm FS}}=2t_{{\rm eq}}\frac{\sinh^{-1}\eta_{{\rm eq}}-\sinh^{-1}\eta_{{\rm dec}}+3/2}{a_{{\rm eq}}\eta_{{\rm eq}}}\thinspace,\label{eq:-42}
\end{equation}
where $\eta\equiv m/p$ with $p$ the momentum of the free-streaming
particle, the subscript ``eq'' denotes the moment of {\it matter-radiation
equality}, and ``dec'' denotes the moment when the particle starts
free-streaming. The matter-radiation equality occurs at $a_{{\rm eq}}\approx1/3420$,
corresponding to a temperature of $T_{{\rm eq}}\approx0.8$ eV. During
free-streaming, the momentum decreases as $p=p_{{\rm dec}}a_{{\rm dec}}/a$,
implying that $\eta_{{\rm eq}}=ma_{{\rm eq}}/(p_{{\rm dec}}a_{{\rm dec}})=\eta_{{\rm dec}}a_{{\rm eq}}/a_{{\rm dec}}$.
The ratio $a_{{\rm eq}}/a_{{\rm dec}}$ can be determined from entropy
conservation, which gives
\begin{equation}
\frac{a_{{\rm dec}}}{a_{{\rm eq}}}=\frac{T_{{\rm eq}}}{T_{{\rm dec}}}\left(\frac{g_{\star{\rm eq}}^{(s)}}{g_{\star{\rm dec}}^{(s)}}\right)^{1/3},\label{eq:-43}
\end{equation}
where $g_{\star}^{(s)}$ is the effective number of relativistic degrees
of freedom in terms of entropy. It is approximately equal to $g_{\star}$
at high temperatures but deviates from $g_{\star}$ after neutrino
decoupling. At $T=T_{{\rm eq}}$, we have $g_{\star}^{(s)}\approx2+7/8\times6\times4/11\approx3.91$
and $g_{\star}\approx2+7/8\times6\times(4/11)^{4/3}\approx3.36$. 

In the limit of large $\eta_{{\rm dec}}$ such that $\sinh^{-1}\eta_{{\rm eq}}\gg\sinh^{-1}\eta_{{\rm dec}}$,
Eq.~\eqref{eq:-42}  actually depends on only three quantities,
$p_{{\rm dec}}/T_{{\rm dec}}$, $g_{\star{\rm dec}}^{(s)}$, and the
mass $m$.  For $p_{{\rm dec}}/T_{{\rm dec}}$, we adopt its thermally-averaged
value, $\langle p/T\rangle_{{\rm dec}}$, which is typically around
three. For instance,  Bose-Einstein and Fermi-Dirac distributions
lead to $\langle p/T\rangle_{{\rm dec}}=2.70$ and $3.15$, respectively~\cite{Coy:2021ann}.

We compute the thermally-averaged value of $\lambda_{{\rm FS}}$ for
our model and compare it with the free-streaming lengths of two well-known
light DM scenarios, namely thermal relic (TR) and sterile neutrino
DM, on which Lyman-$\alpha$ bounds have been reported in several
studies. This allows us to recast the known bounds as a bound on $m_{\phi}$---see
also  Ref.~\cite{Heeck:2017xbu} for a similar approach. Ref.~\cite{Baur:2017stq}
has performed a combined analysis of various Lyman-$\alpha$ data
and reported 4.65 and 28.8 keV as the lower bounds on the masses of
TR and sterile neutrino DM, respectively. The two values correspond
to approximately the same free-streaming length, if  their respective
values of $g_{\star{\rm dec}}^{(s)}$ are used~\cite{Heeck:2017xbu}.
Using Eq.~\eqref{eq:-42}, we obtain $\lambda_{{\rm FS}}\approx0.086$
and $0.087$ Mpc for the two values. In our model, we take $g_{\star{\rm dec}}^{(s)}=106.75$
since $\phi$ as DM is produced at temperatures well above the electroweak
scale, and $\langle p/T\rangle_{{\rm dec}}=2.70$ due to the bosonic
nature. For $\lambda_{{\rm FS}}$ to be below $0.086$ (or $0.087)$
Mpc, the $\phi$ mass needs to be above 13.7 (13.5) keV, respectively.
For simplicity, we take $m_{\phi}\gtrsim14$ keV as the Lyman-$\alpha$
bound on our model. In Fig.~\ref{fig:para-space}, the green shaded
regions are imposed to represent the Lyman-$\alpha$ bound.

In the upper panel of Fig.~\ref{fig:para-space}, we fix $m_{R}$
at $10^{8}$ GeV and plot the blue dashed and dotted lines by requiring
that $\Omega_{\phi}^{(1)}h^{2}$ in Eq.~\eqref{eq:-15} and $\Omega_{\phi}^{(2)}h^{2}$
in Eq.~\eqref{eq:-19} reach the observed DM relic abundance.  The
solid line is obtained by numerically solving the Boltzmann equation,
which will be detailed in Sec.~\ref{sec:Solving}. 

In the lower panels, we present similar plots for $m_{R}=10^{4}$,
$10^{6}$, $10^{10}$, and $10^{14}$ GeV. As is shown here, lighter
$\nu_{R}$ leads to more constrained parameter space to accommodate
a viable DM candidate. The main reason for this is that smaller $m_{R}$
implies larger $\nu_{L}$-$\nu_{R}$ mixing and hence a higher decay
rate of $\phi\to2\nu_{L}$. By requiring that $\Gamma_{\phi\to2\nu_{L}}$
is below $\tau_{{\rm univ}}^{-1}$ or $(11.5\tau_{{\rm univ}})^{-1}$
 and using the lower bound in Eq.~\eqref{eq:th-bound}, we obtain
\begin{equation}
m_{R}\gtrsim\begin{cases}
1.1\times10^{4}\ \text{GeV}\thinspace & \text{for }\Gamma_{\phi\to2\nu_{L}}<\tau_{{\rm univ}}^{-1}\\
5.5\times10^{4}\ \text{GeV}\thinspace & \text{for }\Gamma_{\phi\to2\nu_{L}}<(11.5\tau_{{\rm univ}})^{-1}
\end{cases},\label{eq:-41}
\end{equation}
which can be taken as the lower bound on $m_{R}$ if $\phi$ is to
be considered as DM. If both $\Gamma_{\phi\to2\nu_{L}}<(11.5\tau_{{\rm univ}})^{-1}$
and the Lyman-$\alpha$ bounds are imposed, the lower bound on $m_{R}$
becomes
\begin{equation}
m_{R}\gtrsim2.6\times10^{5}\ \text{GeV}\thinspace.\label{eq:-49}
\end{equation}

\section{Solving the Boltzmann equation\label{sec:Solving}}

Our analytical results in Eqs.~\eqref{eq:-15} and \eqref{eq:-19}
are derived using the freeze-in formalism, which is valid only when
the depletion term in the Boltzmann equation is negligible. For relatively
large $y$ or $y_{\nu}$, the depletion term needs to be taken into
account and one needs to solve the Boltzmann equation to obtain accurate
results. 

To solve the Boltzmann equation numerically, we rewrite Eq.~\eqref{eq:-10}
as follows
\begin{equation}
\frac{d\left(n_{\phi}a^{3}\right)}{da}=\frac{a^{2}}{H}\left(C_{\text{prod}}-C_{\text{depl}}\right).\label{eq:-32}
\end{equation}
For the production term $C_{\text{prod}}$, we include contributions
from the two dominant processes, $2{\cal N}\to2\phi$ and $LH\to{\cal N}\phi$.
The depletion term $C_{\text{depl}}$, caused by the backreactions
$2\phi\to2{\cal N}$ and ${\cal N}\phi\to LH$, would be equal to
$C_{\text{prod}}$ if $\phi$ reaches thermal equilibrium with the
thermal bath. Here we assume that $\phi$ is able to maintain internal
equilibrium with itself such that it has a well-defined temperature
$T_{\phi}$, which can be computed  from $n_{\phi}$. Under this assumption,
$C_{\text{depl}}$ can be computed by replacing $T$ in $C_{\text{prod}}$
with $T_{\phi}$. 

\begin{figure}
\centering

\includegraphics[width=0.7\textwidth]{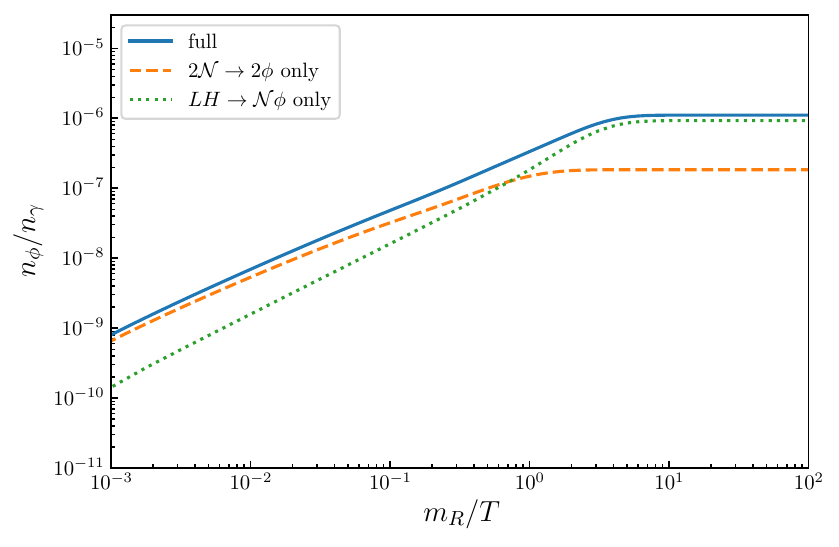}

\caption{Numerical solutions of the Boltzmann equation, with  $m_{R}=10^{8}$
GeV, $m_{\phi}=10^{-4}$ GeV, and $y=3\times10^{-4}$. The solid curve
represent the full numerical solution while the dashed and dotted
lines represents the contributions of $2{\cal N}\to2\phi$ and $LH\to\phi{\cal N}$.
\label{fig:sol}}
\end{figure}

With the above setup, it is straightforward to solve the Boltzmann
equation numerically. Fig.~\ref{fig:sol} shows an example with $m_{R}=10^{8}$
GeV, $m_{\phi}=10^{-4}$ GeV, and $y=3\times10^{-4}$. In addition
to the full numerical solution (blue solid line), we also split it
into two contributions $2{\cal N}\to2\phi$ and $LH\to{\cal N}\phi$,
presented by the dashed and dotted curves there. This example illustrates
the possibility of a scenario where the process $2{\cal N}\to2\phi$
predominates in the initial stage, with $LH\to\phi{\cal N}$ taking
over as the dominant process at a later phase.

As mentioned above, large $y$ would lead to the invalidity of the
freeze-in formalism since $C_{\text{depl}}$ becomes significant.
Indeed, the blue solid curves in Fig.~\ref{fig:para-space} obtained
from numerical solutions exhibit turning points when they approach
 the orange region. Above the turning points, $\Omega_{\phi}h^{2}$
becomes independent of $y$ due to the established equilibrium between
$\phi$ and the thermal bath.

\section{Neutrino and other possible signals from DM \label{sec:Neutrino-signals}}

 As a decaying DM candidate,  the $\nu_{R}$-philic scalar features
interesting neutrino signals for DM indirect detection. Indeed, in
the low-mass regime with $m_{\phi}<2m_{\mu}\approx212$ MeV and $m_{R}<1.5\times10^{10}\ \text{GeV}$
{[}see Eq.~\eqref{eq:-5}{]}, its dominant decay channel is $\phi\to2\nu_{L}$,
implying the  emission of monochromatic neutrino lines, which if detected
would be a smoking-gun signal for DM indirect detection. 

Neutrino signals as a novel approach to DM indirect detection have
been investigated in a number of studies~\cite{Palomares-Ruiz:2007egs,Dudas:2014bca,Garcia-Cely:2017oco,Coy:2020wxp,Coy:2021sse,Hufnagel:2021pso}.
A crucial difference between the neutrino signals in our model and
in various previously considered scenarios is that heavy DM with $m_{\phi}$
above $2m_{\mu}$ in our model  has been ruled out---see Fig.~\ref{fig:para-space}.
Consequently, the neutrino signals are less energetic and the observation
of high-energy astrophysical neutrinos at IceCube, ANTARES, and other
neutrino telescopes cannot be used in our model. At low energies,
searches for extraterrestrial neutrinos have been conducted by Super-Kamiokande
(SK)~\cite{Super-Kamiokande:2002exp,Super-Kamiokande:2013ufi}, Borexino~\cite{Borexino:2010zht},
and KamLAND~\cite{KamLAND:2011bnd},  adopting inverse beta decay
(IBD, $\overline{\nu}_{e}+p\to n+e^{+}$) as the primary detection
channel   due to its relatively low-background, large cross section,
and   successful event identification. 

\begin{figure}
\centering

\includegraphics[width=0.8\textwidth]{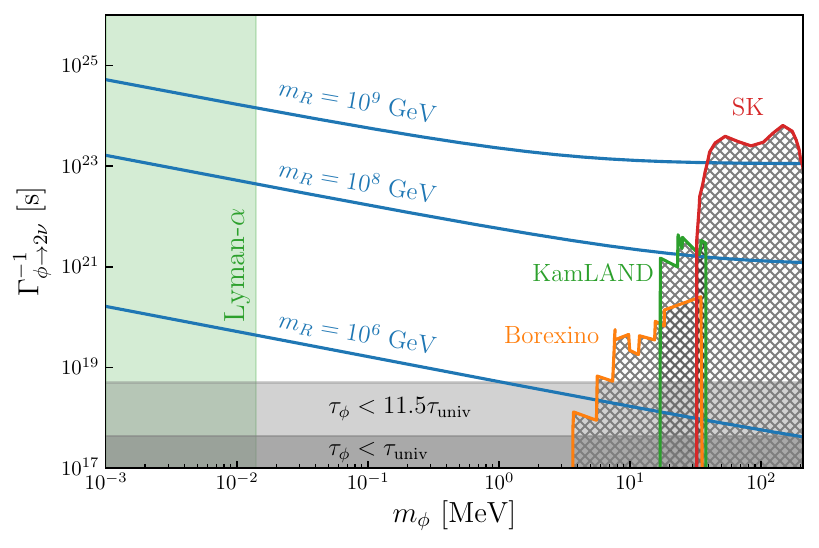}

\caption{The strength of neutrino signals (blue lines) from $\phi$ decay compared
with experimental limits (hatched regions) from SK~\cite{Super-Kamiokande:2002exp,Super-Kamiokande:2013ufi},
Borexino~\cite{Borexino:2010zht}, and KamLAND~\cite{KamLAND:2011bnd}---see
Refs.~\cite{Palomares-Ruiz:2007egs,Garcia-Cely:2017oco} for compilations
of these limits. \label{fig:nu}}
\end{figure}

 In our framework, the decay width of the $\nu_{R}$-philic scalar
DM is determined by $\Omega_{\phi}h^{2}=0.12$,  if $m_{R}$ and
$m_{\phi}$ are fixed at given values. In Fig.~\ref{fig:nu}, we
plot the decay width of $\phi\to2\nu_{L}$ as a function of $m_{\phi}$
for $m_{R}=10^{6}$, $10^{8}$ and $10^{9}$ GeV. This is to be confronted
with the limits reached by the aforementioned neutrino experiments.
 As is shown Fig.~\ref{fig:nu}, for $m_{R}$ ranging from $10^{6}$
to $10^{9}$ GeV, a significant part of the parameter space has already
been excluded by SK, Borexino, and KamLAND. Larger $m_{R}$ leads
to a smaller exclusion interval of $m_{\phi}$. We find that the exclusion
interval vanishes at $m_{R}\approx2.4\times10^{9}$ GeV, implying
that heavier right-handed neutrinos above this mass cannot be probed
by the present neutrino experiments. Future experiments such as Hyper-Kamiokande (HK)~\cite{Abe:2011ts}, DUNE~\cite{DUNE:2015lol}, JUNO~\cite{JUNO:2015zny}, and THEIA~\cite{Theia:2019non}
will in general improve the statistics by about one or two orders
of magnitude.  Correspondingly, the sensitivity reach in terms of
$\Gamma_{\phi\to2\nu}^{-1}$ is expected  to be improved by one or
two orders of magnitude. Since $\Gamma_{\phi\to2\nu}^{-1}$ scales
as $m_{R}^{2}$ for fixed $m_{\nu}$, we expect that $m_{R}\sim10^{10}$
GeV could be probe by future neutrino experiments. 

In Fig.~\ref{fig:nu}, we assume that $\phi$ predominantly decays
to $\overline{\nu}_{e}$, since the experimental searches are conducted
in the IBD channel. It is possible that, by tuning the flavor structure
in the neutrino sector, one obtains a suppressed branching ratio of
$\phi$ to $\overline{\nu}_{e}$. In this case, the experimental sensitivity
would be substantially reduced. We should note here that the flavor
of neutrinos, from production to detection, may be significantly altered
due to neutrino oscillation. In addition, the flavor structure is
not arbitrary and has been partially determined by neutrino oscillation
data. Altogether, it implies that the flavor issue might  play a potentially
important role in interpreting the neutrino signals of our DM model.
However, given the large number of free parameters in the $3\times3$
generalizations of $y$ and $y_{\nu}$, a comprehensive analysis on
the full flavor structure is beyond the scope of this work.  We
leave a more dedicated analysis involving the flavor structure to
future work. 

Finally, let us comment on the direct detection of DM in this model.
The $\nu_{R}$-philic scalar DM interacts with normal matter mainly
via the loop-induced coupling to electrons. So in principle, it could
cause electron recoil signals in direct detection experiments via
elastic scattering, $\phi+e^{-}\to\phi+e^{-}$, or inelastic scattering,
$\phi+e^{-}\to\gamma+e^{-}$. The latter has a much larger cross section
than the former because it is less suppressed by the loop-induced
coupling $y_{\phi ee}$. We have estimated the cross section of $\phi+e^{-}\to\gamma+e^{-}$
and find that it  is more than 20 orders of magnitude below the current
experimental sensitivity.

\section{Conclusion\label{sec:Conclusion}}

In this study, we explore a simple extension of the Type-I seesaw
model, in which a real scalar $\phi$ is introduced and exclusively
coupled to right-handed neutrinos $\nu_{R}$, hence referred to as
the $\nu_{R}$-philic scalar. Our primary objective is to assess whether
the $\nu_{R}$-philic scalar can serve as a viable DM candidate while
preserving the success of Type-I seesaw in explaining the small neutrino
masses. To this end, we systematically investigate the decay channels
of the $\nu_{R}$-philic scalar, compute the thermal production rates
of $\phi$, and, by combining the results regarding these two aspects,
 obtain the viable parameter space for $\phi$ to qualify as DM.

The viable parameter space is presented in Fig.~\ref{fig:para-space}.
Notably, this parameter space necessitates that the scalar mass should
be above $173.4$ eV and below 212 MeV (i.e.~$2m_{\mu}$) for $\phi$
to be DM. Moreover, our research underlines the critical role played
by the mass of $\nu_{R}$ in shaping this parameter space. A lighter
$\nu_{R}$ leads to larger $\nu_{L}$-$\nu_{R}$ mixing and increases
the decay width of $\phi\to2\nu_{L}$, resulting more constrained
parameter space. According to Eq.~\eqref{eq:-41}, the viable parameter
space sets a lower bound of $\sim10^{4}$ GeV for the $\nu_{R}$ mass. 

Further, we consider the observational consequence of the $\nu_{R}$-philic
scalar DM within the viable parameter space. The emission of monochromatic
neutrino lines from $\phi$ decay, if detected, would be a smoking-gun
signal for DM indirect detection. In our model, the neutrino lines
are less energetic than $\sim10^{2}$ MeV and cannot contribute to
the observation of high-energy astrophysical neutrinos at IceCube.
Within the energy range from a few to $\sim10^{2}$ MeV, they could
be detected by the present neutrino detectors.

In summary, our investigation provides a comprehensive exploration
of the $\nu_{R}$-philic scalar as a DM candidate, shedding light
on the interplay between neutrinos and DM. The constraints we have
identified regarding the masses of $\phi$ and $\nu_{R}$ constitute
a valuable foundation for forthcoming experimental and theoretical
explorations aimed at deciphering the enigmatic nature of DM and its
connections to a broader landscape of particle physics.

\appendix

\section{Squared amplitudes,  cross sections, and collision terms\label{sec:M-and-C}}

In this appendix, we present detailed calculations of squared amplitudes,
 cross sections, and collision terms used in this work.   Regarding
the Majorana feature of $\nu_{R}$, although the two-component notation
of Weyl spinors is conceptually simple, it is more convenient to adopt
four-component spinors due to well-developed computing packages. Hence
we rewrite the $\nu_{R}\nu_{R}\phi$ term in Eq.~\eqref{eq:L} in
terms of Majorana spinors:
\begin{equation}
\mathcal{L}\supset\frac{y}{2}\overline{\Psi}_{{\cal N}}\phi\Psi_{{\cal N}}\thinspace,\label{eq:-24}
\end{equation}
where $\Psi_{{\cal N}}$ is given by Eq.~\eqref{eq:-47}. Note that,
unlike Eq.~\eqref{eq:L}, here we do not add an ``h.c.'' term because
$y$ is real. If $y$ were  complex, then the Lagrangian would  be
$\mathcal{L}\supset\frac{1}{2}\overline{\Psi}_{{\cal N}}(yP_{L}+y^{*}P_{R})\phi\Psi_{{\cal N}}$.

\subsection{$2{\cal N}\rightarrow2\phi$}

Since $\phi$ is a real scalar, there are two diagrams contributing
to this process, the $t$-channel and $u$-channel diagrams. Including
the two channels, the scattering amplitude of $2{\cal N}\rightarrow2\phi$
 reads:
\begin{equation}
i{\cal M}_{2{\cal N}\rightarrow2\phi}=-iy^{2}\overline{v}_{2}\frac{\slashed{q}_{1}+m_{R}}{t-m_{R}^{2}}u_{1}-iy^{2}\overline{v}_{2}\frac{\slashed{q}_{2}+m_{R}}{u-m_{R}^{2}}u_{1}\thinspace,\label{eq:-25}
\end{equation}
where $\overline{\nu}_{2}$ and $u_{1}$ denote the two initial state
of ${\cal N}$, and $q_{1}(q_{2})$ is the momentum of the $t$($u$)-channel
propagator. Hence, $t=q_{1}^{2}$ and $u=q_{2}^{2}$. After applying
the standard procedure of computing traces, we get
\begin{align}
|{\cal M}_{2{\cal N}\rightarrow2\phi}|^{2} & =2y^{4}\frac{tu\left(t-u\right)^{2}-m_{R}^{2}\left(t+u\right)\left(t^{2}+14tu+u^{2}\right)}{\left(t-m_{R}^{2}\right)^{2}\left(u-m_{R}^{2}\right)^{2}}\nonumber \\
 & \phantom{=}+2y^{4}\frac{m_{R}^{4}\left(t^{2}+30tu+u^{2}\right)+16m_{R}^{6}\left(t+u\right)-32m_{R}^{8}}{\left(t-m_{R}^{2}\right)^{2}\left(u-m_{R}^{2}\right)^{2}}\thinspace.\label{eq:-26}
\end{align}

The cross section of $2{\cal N}\rightarrow2\phi$ in the center-of-mass
frame is
\begin{equation}
\begin{aligned}\sigma_{2{\cal N}\rightarrow2\phi} & =\frac{1}{S}\int\frac{\beta_{f}|{\cal M}_{2{\cal N}\rightarrow2\phi}|^{2}}{64\pi^{2}s\beta_{i}}d\Omega\\
 & =\frac{y^{4}}{4\pi s\beta_{i}^{2}}\left(2\beta_{i}^{3}-3\beta_{i}+\left(3-2\beta_{i}^{4}\right)\tanh^{-1}\beta_{i}\right),
\end{aligned}
\end{equation}
where $\beta_{i}$ and $\beta_{f}$ denote the velocities of the initial
and final particles, and $S=2$ is the symmetry factor due to identical
particles. In the above calculation, we have assumed $m_{R}\gg m_{\phi}$
so that $\beta_{f}\approx1$. 

The collision term can be computed by further integrating the cross
section~\cite{Gondolo:1990dk}:

\begin{equation}
C_{2{\cal N}\rightarrow2\phi}\approx\frac{TN_{f}}{32\pi^{4}}\int_{4m_{R}^{2}}^{\infty}s^{\frac{1}{2}}(s-4m_{R}^{2})\sigma_{2{\cal N}\rightarrow2\phi}K_{1}\left(\frac{\sqrt{s}}{T}\right)ds\thinspace.\label{eq:ct}
\end{equation}
Analytically, the integral can only be computed in the low-$T$ ($T\ll m_{R}$)
and high-$T$ ($T\gg m_{R}$) limits. 

In the low-$T$ limit, $\beta_{i}\rightarrow0$, we can expand $\tanh^{-1}\beta_{i}$
in terms of polynomial of $\beta_{i}$. The leading order gives 
\begin{equation}
\sigma_{2{\cal N}\rightarrow2\phi}\approx\frac{3y^{4}\beta_{i}}{4\pi s}\thinspace.\label{eq:-27}
\end{equation}
Substituting Eq.~\eqref{eq:-27} into Eq.~\eqref{eq:ct}, we obtain
\begin{equation}
C_{2{\cal N}\rightarrow2\phi}\approx\frac{9y^{4}N_{f}}{128\pi^{4}}T^{4}e^{-2m_{R}/T}\thinspace.\label{eq:-28}
\end{equation}

As for the high-$T$ limit, we can expand the cross section $\sigma_{2{\cal N}\rightarrow2\phi}$
in terms of $m_{R}$. Neglecting terms of ${\cal O}(m_{R}^{2})$,
we obtain
\begin{equation}
\sigma_{2{\cal N}\rightarrow2\phi}\approx\frac{y^{4}}{8\pi s}\left[\ln\left(\frac{s}{m_{R}^{2}}\right)-2\right].\label{eq:-29}
\end{equation}
Substituting Eq.~\eqref{eq:-29} into Eq.~\eqref{eq:ct} and assuming
$\ln(T/m_{R})\gg1$, we obtain
\begin{equation}
C_{2{\cal N}\rightarrow2\phi}\approx\frac{y^{4}N_{f}}{32\pi^{5}}T^{4}\ln\left(\frac{2T}{m_{R}}\right).\label{eq:-30}
\end{equation}

Based on Eqs.~\eqref{eq:-28} and \eqref{eq:-30}, one can construct
an analytical expression that approaches Eqs.~\eqref{eq:-28} and
\eqref{eq:-30} in the low-$T$ and high-$T$ limits. The expression
is presented in Eq.~\eqref{eq:-8}. 

\begin{figure}
\centering

\includegraphics[width=0.6\textwidth]{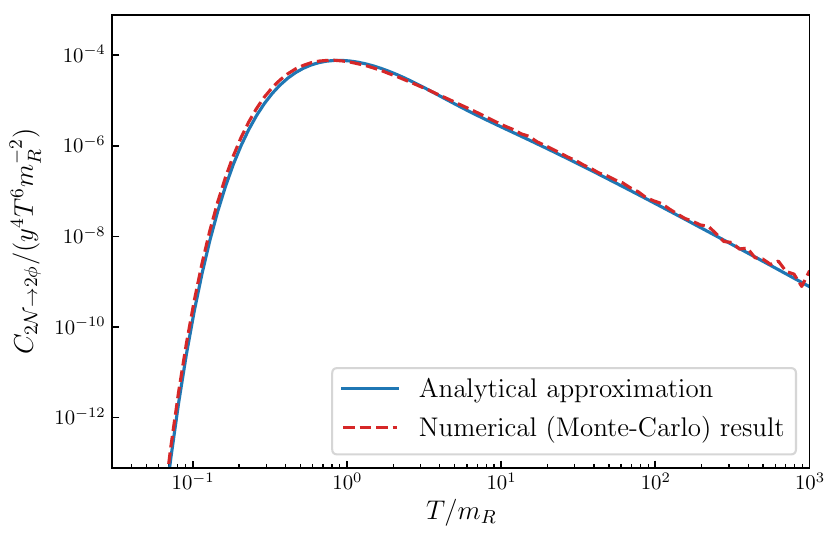}

\caption{Analytical approximation of the collision term $C_{2{\cal N}\to2\phi}$
{[}given in Eq.~\eqref{eq:-8}{]} compared with the numerical result
obtained using the Monte-Carlo method. \label{fig:monte}}
\end{figure}

Let us compare our analytical result with the numerical result obtained
by directly evaluating the following integral:
\begin{equation}
C_{2{\cal N}\to2\phi}=N_{f}\int d\Pi_{1}d\Pi_{2}d\Pi_{3}d\Pi_{4}f_{{\cal N}}^{2}|{\cal M}_{2{\cal N}\rightarrow2\phi}|^{2}(2\pi)^{4}\delta^{4}\thinspace,\label{eq:-31}
\end{equation}
where $d\Pi_{i}\equiv\frac{d^{3}\mathbf{p}_{i}}{2E_{i}(2\pi)^{3}}$
with $i$ indicating the $i$-th particles in the process ${\cal N}+{\cal N}\rightarrow\phi+\phi$.
The momentum distribution function $f_{{\cal N}}$ takes the Fermi-Dirac
distribution. We adopt the Monte-Carlo method introduced in Appendix~B
of Ref.~\cite{Luo:2020fdt} to perform the integration. Fig.~\ref{fig:monte}
shows the comparison of the analytical expression in Eq.~\eqref{eq:-8}
with the numerical result. Obviously, Eq.~\eqref{eq:-8} is quite
accurate when compared with  the numerical result. 

\subsection{$LH\rightarrow{\cal N}\phi$}

 In the main text of this paper, $LH\rightarrow{\cal N}\phi$ actually
refers to two processes, $LH\rightarrow{\cal N}\phi$ and its antiparticle
companion, $\overline{L}\overline{H}\rightarrow{\cal N}\phi$. 

For $LH\rightarrow{\cal N}\phi$, the scattering amplitude reads:
\[
i{\cal M}_{LH\rightarrow{\cal N}\phi}=-i\frac{yy_{\nu}}{s-m_{R}^{2}}\overline{u}_{3}\left(\slashed{p}+m_{R}\right)P_{L}u_{1}\thinspace,
\]
where $\overline{u}_{3}$ and $u_{1}$ denote the final state of ${\cal N}$
and the initial state of $L$, respectively. The momentum of the $s$-channel
propagator in this process is denoted by $p$, with $s=p^{2}$.  

The squared amplitude is then given by

\begin{equation}
\begin{aligned}|{\cal M}_{LH\rightarrow{\cal N}\phi}|^{2} & =\frac{y^{2}y_{\nu}^{2}}{\left(s-m_{R}^{2}\right)^{2}}\text{Tr}\left[\left(\slashed{p}_{3}+m_{R}\right)\left(\slashed{p}+m_{R}\right)P_{L}\slashed{p}_{1}P_{R}\left(\slashed{p}+m_{R}\right)\right]\\
 & =\frac{y^{2}y_{\nu}^{2}\left(m_{R}^{4}+m_{R}^{2}\left(3s-t\right)-su\right)}{\left(s-m_{R}^{2}\right)^{2}}\thinspace,
\end{aligned}
\end{equation}
where we have neglected all particle masses except for $m_{R}$. 

Note that $L$ and $H$ are electroweak $SU(2)$ doublets. In this
regard, we interpret the above result as the squared amplitude of
each component of $L$ scattering off the corresponding component
of $H$. For cross sections presented below, we also adopt the same
convention. The combined contribution of the two $SU(2)$ components
in $L$ (i.e. $\nu_{L}$ and $e_{L}$), together with the contribution
of the antiparticle channel ($\overline{L}\overline{H}\rightarrow{\cal N}\phi$),
will be summed in the final result of the collision term. 

Integrating the squared amplitude, we obtain the cross section
\begin{equation}
\sigma_{LH\rightarrow{\cal N}\phi}=\frac{y^{2}y_{\nu}^{2}}{32\pi s^{2}}\cdot\frac{m_{R}^{4}+6m_{R}^{2}s+s^{2}}{s-m_{R}^{2}}\thinspace.\label{eq:-35}
\end{equation}

For $\overline{L}\overline{H}\rightarrow{\cal N}\phi$, the scattering
amplitude reads:
\begin{equation}
i{\cal M}_{\overline{L}\overline{H}\rightarrow{\cal N}\phi}=-i\frac{yy_{\nu}}{s-m_{R}^{2}}\overline{v}_{1}\left(\slashed{p}+m_{R}\right)P_{R}v_{3}\thinspace,\label{eq:-33}
\end{equation}
where $\overline{v}_{1}$ and $v_{3}$ denote the initial state of
$L$ and the final state of ${\cal N}$, respectively. 

Correspondingly, the squared amplitude is

\begin{equation}
\begin{aligned}|{\cal M}_{\overline{L}\overline{H}\rightarrow{\cal N}\phi}|^{2} & =\frac{y^{2}y_{\nu}^{2}}{\left(s-m_{R}^{2}\right)^{2}}\text{Tr}\left[\slashed{p}_{1}\left(\slashed{p}+m_{R}\right)P_{R}\left(\slashed{p}_{3}-m_{R}\right)P_{L}\left(\slashed{p}+m_{R}\right)\right]\\
 & =\frac{yy_{\nu}^{2}\left(m_{R}^{4}+m_{R}^{2}\left(s-t\right)-su\right)}{\left(s-m_{R}^{2}\right)^{2}}\thinspace,
\end{aligned}
\end{equation}
which leads to the following cross section
\begin{equation}
\sigma_{\overline{L}\overline{H}\rightarrow{\cal N}\phi}=\frac{y^{2}y_{\nu}^{2}}{32\pi s^{2}}\cdot\frac{\left(s+m_{R}^{2}\right)^{2}}{s-m_{R}^{2}}.\label{eq:-34}
\end{equation}

We see that in the high-energy limit, $s\gg m_{R}$, Eqs.~\eqref{eq:-35}
and \eqref{eq:-34} lead to the same result
\begin{equation}
\lim_{s\to\infty}\sigma_{LH\rightarrow{\cal N}\phi}=\lim_{s\to\infty}\sigma_{\overline{L}\overline{H}\rightarrow{\cal N}\phi}=\frac{y^{2}y_{\nu}^{2}}{32\pi s}\thinspace,\label{eq:-36}
\end{equation}
while in the low-energy limit, $s-m_{R}^{2}\equiv\delta\to0$, Eqs.~\eqref{eq:-35}
and \eqref{eq:-34} differ by a factor of two: 
\begin{equation}
\lim_{s\to m_{R}^{2}}\sigma_{LH\rightarrow{\cal N}\phi}=\frac{y^{2}y_{\nu}^{2}}{4\pi\delta}\thinspace,\ \ \lim_{s\to m_{R}^{2}}\sigma_{\overline{L}\overline{H}\rightarrow{\cal N}\phi}=\frac{y^{2}y_{\nu}^{2}}{8\pi\delta}\thinspace.\label{eq:-37}
\end{equation}

The collision term responsible for the production of $\phi$ via the
above two processes, including the independent contributions from
the $\nu_{L}$ and $e_{L}$ components in $L$, is computed by
\begin{equation}
C_{LH\rightarrow{\cal N}\phi}\approx\frac{TN_{f}}{32\pi^{4}}\int_{m_{R}^{2}}^{\infty}\left(2\sigma_{LH\rightarrow{\cal N}\phi}+2\sigma_{\overline{L}\overline{H}\rightarrow{\cal N}\phi}\right)s^{1/2}\left(s-m_{R}^{2}\right)K_{1}\left(\frac{\sqrt{s}}{T}\right)ds\thinspace.\label{eq:ct-1}
\end{equation}
At high temperatures, we substitute Eq.~\eqref{eq:-36} into Eq.~\eqref{eq:ct-1}
and obtain
\begin{align}
C_{LH\rightarrow{\cal N}\phi} & \approx\frac{y^{2}y_{\nu}^{2}N_{f}}{64\pi^{5}}m_{R}T^{3}K_{1}\left(\frac{m_{R}}{T}\right)\label{eq:-38}\\
 & \approx\frac{y^{2}y_{\nu}^{2}N_{f}}{64\pi^{5}}T^{4}\thinspace,\ \ \ (T\gg m_{R})\thinspace.\label{eq:-39}
\end{align}
At low temperatures, we substitute Eq.~\eqref{eq:-37} into Eq.~\eqref{eq:ct-1}
and obtain
\begin{align}
C_{LH\rightarrow{\cal N}\phi} & \approx\frac{y^{2}y_{\nu}^{2}N_{f}}{64\pi^{5}}3m_{R}^{2}T^{2}K_{2}\left(\frac{m_{R}}{T}\right),\ \ \ \ \ \ (T\ll m_{R})\thinspace.\label{eq:-40}
\end{align}
Since $\lim_{x\to0}K_{1}\left(x\right)/K_{2}\left(x\right)=x/2$,
we propose the following expression that can cover both the high-$T$
and low-$T$ limits:
\begin{align}
C_{LH\rightarrow{\cal N}\phi} & \approx\frac{y^{2}y_{\nu}^{2}N_{f}}{64\pi^{5}}3m_{R}^{2}T^{2}K_{2}\left(\frac{m_{R}}{T}\right)\frac{m_{R}^{2}+T^{2}}{m_{R}^{2}+6T^{2}}\thinspace.\label{eq:-40-1}
\end{align}
As we have checked, when compared to numerical results from the Monte-Carlo
integration, Eq.~\eqref{eq:-40-1} also exhibits excellent accuracy. 

\begin{acknowledgments}
This work is supported in part by the National Natural Science Foundation
of China under grant No.~12141501 and also supported by CAS Project
for Young Scientists in Basic Research (YSBR-099). 
\end{acknowledgments}

\bibliographystyle{JHEP}
\bibliography{ref}

\end{document}